\documentclass{iaus}
\usepackage{graphicx}
  \checkfont{eurm10}
  \iffontfound
    \IfFileExists{upmath.sty}
      {\typeout{^^JFound AMS Euler Roman fonts on the system,
                   using the 'upmath' package.^^J}%
       \usepackage{upmath}}
      {\typeout{^^JFound AMS Euler Roman fonts on the system, but you
                   dont seem to have the}%
       \typeout{'upmath' package installed. iaus.cls can take advantage
                 of these fonts,^^Jif you use 'upmath' package.^^J}%
      }
  \else
  \fi

  \checkfont{msam10}
  \iffontfound
    \IfFileExists{amssymb.sty}
      {\typeout{^^JFound AMS Symbol fonts on the system, using the
                'amssymb' package.^^J}%
       \usepackage{amssymb}%
         \let\leq=\leqslant
         
      }{}
  \fi

  \IfFileExists{amsbsy.sty}
    {\typeout{^^JFound the 'amsbsy' package on the system, using it.^^J}%
     \usepackage{amsbsy}}
    {\providecommand\boldsymbol[1]{\mbox{\boldmath $##1$}}}

\newsavebox{\astrutbox}
\sbox{\astrutbox}{\rule[-5pt]{0pt}{20pt}}

\title[The Interplay among Black Holes, Stars and ISM in Galactic 
       Nuclei]{QSO Formation under Coevolution of SMBH and Bulge}

\author[N.Kawakatu{\it et al.\/}]%
{N.Kawakatu$^1$, %
M.Umemura $^1$\break
\and M.Mori $^2$}

\affiliation{$^1$ Center for Computational Physics, University of 
Tsukuba, Ten-nodai, 1-1-1, Ibaraki 305-8577, Japan
 email: kawakatu@rccp.tsukuba.ac.jp, umemura@rccp.tsukuba.ac.jp\\[\affilskip]
$^2$ Department of Law, Senshu University, Tama-Ku,
Kawasaki 214-8580, Japan email: mmori@isc.senshu-u.ac.jp}

\pubyear{2004}
\volume{222}
\pagerange{1--8}
\date{?? and in revised form ??}
\jname{The Interplay among Black Holes, Stars and ISM \\in Galactic Nuclei}
\editors{Th. Storchi Bergmann, L.C. Ho \& H.R. Schmitt, eds.}
\begin{document}

\maketitle

\begin{abstract}

The formation and growth of supermassive black holes
(SMBHs) physically linked with bulges are considered. 
We focus on the radiation hydrodynamic process for the 
growth of SMBH in the optically thick starburst phase, 
where radiation from bulge stars drives the mass
accretion on to a galactic center through radiation drag effect.
In the present scenario, the AGN luminosity-dominant
phase (QSO phase) is preceded by the host
luminosity-dominat phase, which is called ``proto-QSO
phase''. In this phase, there exists the massive dusty
disks within younger bulges. 
Also, the proto-QSO phase is anticipated by an
optically-thick ultraluminous infrared galaxy (ULIRG) 
phase.
Furthermore, such radiation hydrodynamic model has been 
also applied to disk galaxies. It turns out that the
mass of a SMBH primarily correlates with a bulge
component even in a disk galaxy.
Thus, by analogy to proto-QSOs, the BH growing phase 
in disk galaxies may have massive dusty disks 
within younger bulges.
\end{abstract}

\firstsection 
\section{Introduction}

The paradigm that ultraluminous infrared galaxies
(ULIRGs) could evolve into QSOs was proposed by
pioneering studies by Sanders et al. (1988) and 
Norman \& Scovill (1988).
By recent observations, the X-ray emission (Brandt et
a. 1997) or Pa$\alpha$ lines (Veilleux, Sanders, \& Kim
1999) intrinsic for active galactic nuclei (AGNs) have
been detected in more than one third of ULIRGs. 
On the other hand, recent high-resolution observations of galactic centers have
revealed that the estimated mass of a central ``massive
dark object''(MDO), which is the nomenclature for a
supermassive BH candidate, does correlate with the mass
of a galactic bulge; the mass ratio of the BH to the
bulge is 0.002 as a median value (e.g., Marconi \& Hunt 2003). 
In addition, it has been found that QSO host galaxies
are mostly luminous and well-evolved early-type galaxies 
(e.g., McLure, Dunlope, \& Kukula 2000). 
Comprehensively judging from all these findings, 
it is likely that ULIRGs, QSOs, Bulges, and SMBHs are
physically related to each other.

\section{Formation of Supermassive Black Holes}\label{}

A radiation drag model for the formation of SMBHs is
recently proposed by Umemura (2001).
Here, we suppose a simple two-component system that consists of a spheroidal stellar bulge 
and inhomogeneous optically-thick interstellar medium (ISM) within it.
In this model, radiation drag extracts the angular
momentum from inhomogeneous optically-thick ISM and
allow it to accrete onto the center. 
Then, the mass of an MDO, $M_{\rm MDO}$, 
which is the total mass of dusty ISM assembled to the central massive 
object, is given by 

\begin{equation}
M_{\rm MDO}=\eta_{\rm drag}\int^{t_{\rm w}}_{t_{\rm thin}}\frac{L_{\rm bulge}(t)}{c^2}dt,
\end{equation}
where $L_{\rm bulge}$ is the bulge luminosity, $t_{\rm w}$ is a
galactic wind timescale, 
and $t_{\rm thin}$ is a time before which the optical
depth is less than unity.
Here, $\eta_{\rm drag}$ is found to be maximally 
0.34 in the optically thick limit based on the numerical
simulation by Kawakatu \& Umemura (2002).

In this paper, we should distinguish BH mass from the mass of an MDO 
although the mass of an MDO is often regarded as BH mass from an observational point of view.
Supposing the mass accretion driven by the viscosity on to the BH horizon 
is limited by an order of Eddington rate, 
the BH mass grows according to 

\begin{equation}
M_{\rm BH}=M_{0}e^{\nu t/t_{\rm Edd}},
\label{eq2}
\end{equation} 
where $\nu$ is the ratio of BH accretion rate to the Eddington rate,  
and $t_{\rm Edd}$ is the Eddington timescale, $t_{\rm Edd}=1.9\times 10^{8}{\rm yr}$.
Here $M_{0}$ is the mass of a seed BH, 
which could be a massive BH with $\sim 10^{5}M_{\odot}$
formed by the collapse of a rotating supermassive
star (Shibata \& Shapiro 2002).

\begin{figure}
\begin{center}
\includegraphics[height=70mm,clip]{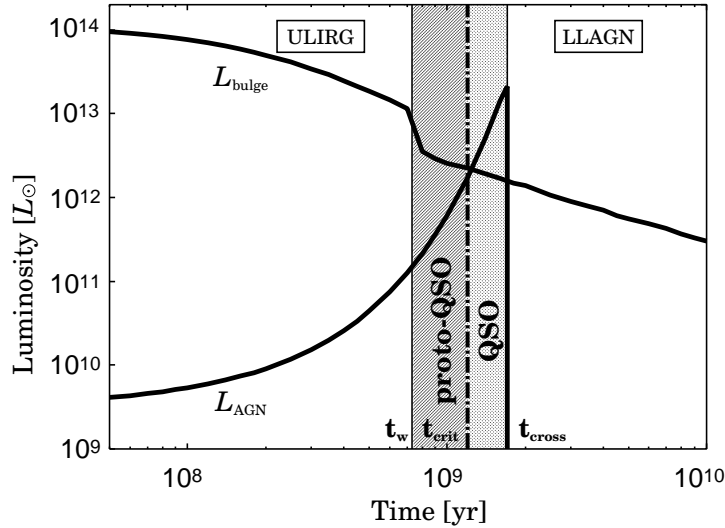}
\end{center}
\caption{
{\scriptsize 
AGN and bulge luminosity as a function of time.
The ordinate is the luminosity in units of $L_{\odot}$.
$t_{\rm crit}$ is the time when $L_{\rm bulge}=L_{\rm AGN}$.
Here, we assume that $L_{\rm AGN}$ is the Eddington luminosity.
The phase at $t<t_{\rm w}$ is a bright and optically thick phase,
which may correspond to a ultraluminous infrared galaxy (ULIRG) phase.
After the AGN luminosity ($L_{\rm AGN}$) exhibits a peak at $t_{\rm cross}$, 
it fades out abruptly.
The later fading nucleus could be a low luminosity AGN (LLAGN).
The optically-thin, bright AGN phase ({\it gray area}) can be divided into two phases; 
one is the host-dominant phase (proto-QSO), which is the dark gray area 
($t_{\rm w}\leq t \leq t_{\rm crit}$) and the other is the AGN-dominant phase
(QSO), 
which is the light gray area ($t_{\rm crit}\leq t \leq t_{\rm cross}$).
The lifetime of both phases are comparable, $\approx
10^{8}$yr.
}
}
\label{fig:1}
\end{figure}

\section{Coevolution of BH Growth and Bulge}

Here, we construct a scenario of the coevolution of SMBH
and bulge based on the radiation drag model for SMBH
formation. 
In order to treat the realistic chemical evolution of
host galaxy, we use an evolutionary spectral synthesis 
code 'PEGASE'(Fioc \& Rocca-Volmerange 1997). 
Also, we employ a galactic wind model with the wind
epoch of $t_{\rm w}=7\times 10^{8}$yr 
because it can reproduce a present-day color-magnitude
relation. In this model, the system is assumed to change
from optically-thick to optically-thin phase at $t_{\rm
w}$. 
Also, we assume the star formation rate is in proportion to
gas fraction and initial gas mass is $10^{12}M_{\odot}$.
Thereby, we can estimate the evolution of the physical
properties of QSO host, such as 
mass, luminosity, color and metallicity.

Based on the present coevolution model, the mass
accretion proportional to the bulge luminosity leads to the growth 
of an MDO, which is likely to form a massive dusty disk 
in the nucleus. However, the matter in the MDO does not promptly fall into the BH, 
because the BH accretion is limited by equation ~(\ref{eq2}). 
The BH mass reaches $M_{\rm MDO}$ at a time $t_{\rm
cross}$ because almost all of the MDO matter has fallen
onto the central BH.  
The resultant BH fraction becomes $f_{\rm BH}\simeq 0.001$, 
which is just comparable to the observed ratio.
The evolution of bulge luminosity ($L_{\rm bulge}$) and 
AGN luminosity ($L_{\rm AGN}$) are shown in
Figure~\ref{fig:1}, assuming the constant Eddington
ratio ($\nu=1$).
Even after the galactic wind ($t>t_{\rm w}$), $M_{\rm
BH}$ continues to grow until $t_{\rm cross}$ and
therefore the AGN brightens with time. 
After $L_{\rm AGN}$ exhibits a peak at $t_{\rm cross}$, 
it fades out abruptly to exhaust the fuel. The fading nucleus could
be a low luminosity AGN (LLAGN).

\begin{figure}
\begin{center}
\includegraphics[height=80mm,clip]{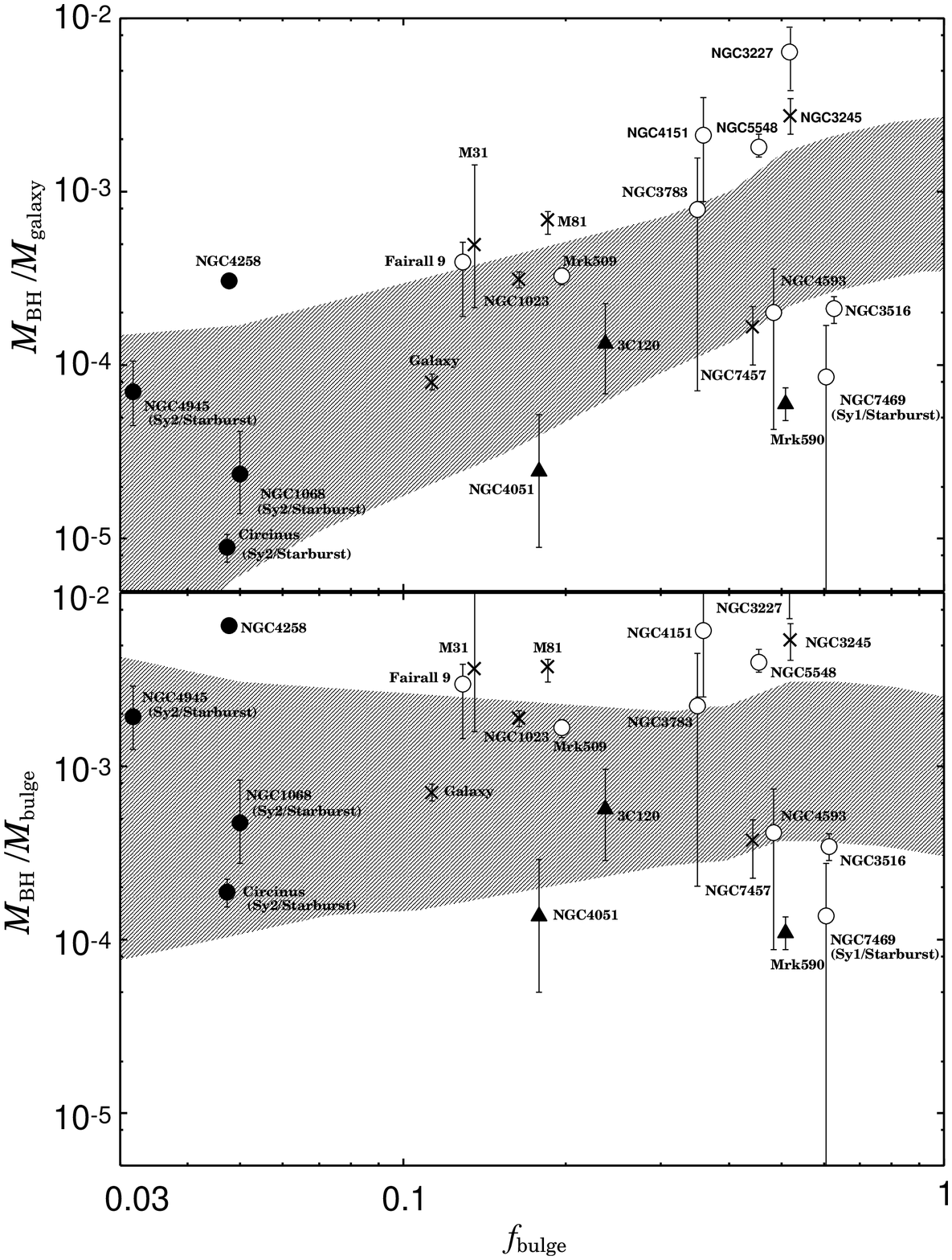}
\end{center}
\caption{
{\scriptsize
Comparison between the theoretical prediction and the observational data
on BH mass for disk galaxies.
The upper panel shows the BH-to-galaxy mass ratio 
($M_{\rm BH}/M_{\rm galaxy}$) and
the lower panel shows the BH-to-bulge mass ratio 
($M_{\rm BH}/M_{\rm bulge}$) against 
the bulge fraction ($M_{\rm bulge}/M_{\rm galaxy}$).
The hatched area is the prediction of the present analysis.
The observational data are plotted by symbols. 
The data points are categorized into four types.
{\it Crosses} -- disk galaxies which do not possess AGNs,
{\it open circles} -- Seyfert 1 galaxies (Sy1s),
{\it filled triangles} -- narrow line Seyfert 1 galaxies (NLSy1s),
and {\it filled circles} -- Seyfert 2 galaxies (Sy2s).
Seyfert galaxies accompanied by starburst activities are specified like
Sy1/starburst or Sy2/starburst.
}
}
\label{fig:2}
\end{figure}

It is found that the area of $t_{\rm w} < t <t_{\rm cross}$ 
can be divided into two phases with a transition time 
$t_{\rm crit}$ when $L_{\rm bulge}=L_{\rm AGN}$; 
the earlier phase is the host luminosity-dominant phase, 
and the later phase is the AGN luminosity-dominant phase.
Also, lifetimes of both phases are comparable to each other, 
which is about $10^{8}$yr.
The AGN-dominant phase is likely to correspond to ordinary QSOs, 
but host-dominant phase is obviously different from observed QSOs so far.
We define this phase as ``a proto-QSO'' (Kawakatu,
Umemura, \& Mori 2003).
We have predicted the observable properties of proto-QSOs as follows:
(1) The broad emission lines are narrower, which
is less than $1500$km/s. Thus, proto-QSO can be regarded
as a ``Narrow line Type I QSO (NLQSO)''
(2) A massive dusty disk ($ > 10^{8}M_{\odot}$)
surrounds a massive BH, and it may obscure the nucleus in the edge-on view to
form a type 2 nucleus.
(3) The BH-to-bulge mass ratio, $M_{\rm BH}/M_{\rm
bulge}$, rapidly increases from $10^{-5.3}$ to
$10^{-3.9}$ in $\approx 10^{8}$yr.
(4) The colors of $({\it V-K})$ at observed bands 
are about 0.5 magnitude bluer than those of QSOs.
The predicted properties of proto-QSOs are similar to
those of high redshift radio galaxies.

The proto-QSO phase is preceded by an optically thick
phase before the galactic wind, which may correspond to
ULIRGs. The present model predicts that the BH fraction 
is anticipated to be much less than 0.002 and grows with 
metallicity in the ULIRG phase.   

\section{Growth of SMBH in Disk Galaxies}

We have hitherto applied the radiation drag model to
elliptical galaxies, but it could been also applied to disk
galaxies. Thus, we recently elucidate the efficiency
of the radiation drag in disk galaxies (Kawakatu \&
Umemura 2004).
As seen in Figure~\ref{fig:2}, it is found that the SMBH should be smaller 
in a disk galaxy, but correlate with the
bulge component in a similar way to an elliptical galaxy.
In addition, the observational trends in disk galaxies are broadly
consistent with the theoretical prediction.
Hence, by analogy to proto-QSOs, a growing BH phase 
in a disk galaxy (e.g., NLS1) possibly have a massive dusty
disk within a younger bulge.

\section{Conclusions}

Based on the radiation drag model for the BH growth,
incorporating the chemical evolution of the early-type
host galaxy, we have built up the coevolution model for 
a QSO BH and the host galaxy. 
As a consequence, we have shown the possibility of the
proto-QSO phase, which is optically thin and host
luminosity-dominant, and has the lifetime comparable to
the QSO phase timescale of a few $10^{8}$ yr. 
Also, by considering theoretical predictions for the
observable properties in proto-QSO phase, we conclude
that radio galaxies at high redshifts are a possible
candidate for proto-QSOs. 
The proto-QSO phase is preceded by an optically-thick
ultraluminous infrared galaxy (ULIRG) phase.
Furthermore, the present model could be applied to disk
galaxies. We found that the mass of a SMBH correlates
with that of a bulge even in disk galaxies.
Thus, by analogy to proto-QSOs, a growing BH phase in a
disk galaxy (e.g., NLS1) may have a massive dusty
disk within a younger bulge. 
In summary, the present model could be a physical
picture of evolution of ULIRGs (LIRGs) to QSOs (Sy1s).


\end{document}